\documentclass[twocolumn,prl,amsmath,amssymb,superscriptaddress,groupedaddress,showpacs]{revtex4}

\usepackage{epsfig}
\usepackage{dcolumn}
\usepackage{bm}
\usepackage{amssymb}

\begin{document}

\title{Transport Coefficients of Hadronic Matter near $T_c$}

\author{Jacquelyn Noronha-Hostler}
\affiliation{Frankfurt Institute for Advanced Studies, Frankfurt, Germany}
\affiliation{Department of
Physics, Columbia University, 538 West 120$^{th}$ Street, New York,
NY 10027, USA}
\affiliation{Institut f\"ur Theoretische Physik, Goethe Universit\"at, Frankfurt, Germany}
\author{Jorge Noronha}
\affiliation{Department of
Physics, Columbia University, 538 West 120$^{th}$ Street, New York,
NY 10027, USA}
\author{Carsten Greiner}
\affiliation{Institut f\"ur Theoretische Physik, Goethe Universit\"at, Frankfurt, Germany}

\begin{abstract}

A hadron resonance gas model including all known particles and resonances with masses $m<2$ GeV and an exponentially rising density of Hagedorn states for $m>2$ GeV is used to obtain an upper bound on the shear viscosity to entropy density ratio, $\eta/s\approx1/(4\pi)$, of hadronic matter near $T_c$. We found a large trace anomaly and small speed of sound near $T_c$, which agree well with recent lattice calculations. We comment on the bulk viscosity to entropy density ratio close to $T_c$.
\end{abstract}
\pacs{25.75.Nq, 51.20.+d}
\maketitle


The large azimuthal asymmetry of low-$p_T$ particles and the strong quenching of high-$p_T$ probes measured at RHIC \cite{whitepapers} indicate that the new state of matter produced in heavy ion collisions is a strongly interacting quark-gluon plasma \cite{Gyulassy:2004zy}. The matter formed in these collisions behaves almost as a perfect liquid \cite{etas} characterized by a very small value for its shear viscosity to entropy density ratio, which is in the ballpark of the lower bound $\eta/s \geq 1/(4\pi)$ \cite{Policastro:2001yc} derived within the anti-de Sitter/conformal field theory (AdS/CFT) correspondence \cite{Maldacena:1997re}. It was further conjectured by Kovtun, Son, and Starinets (KSS) \cite{Kovtun:2004de} that this bound holds for all substances in nature (see, however, Refs.\ \cite{Cohen:2007qr,Dobado:2007tm} for possible counterexamples involving nonrelativistic systems).

Recent lattice calculations \cite{Meyer:2007ic} in pure glue $SU(3)$ gauge theory have shown that $\eta/s$ remains close to the KSS bound at temperatures not much larger than $T_c$. Additionally, calculations within the BAMPS parton cascade \cite{xucarsten}, which includes inelastic gluonic $gg\leftrightarrow ggg$ reactions, showed that $\eta/s \sim 0.13$ in a pure gluon gas \cite{Xu:2007ns}. Moreover, it was argued in \cite{Csernai:2006zz} that this ratio should have a minimum at (or near) the phase transition in quantum chromodynamics (QCD). This is expected because $\eta/s$ increases with decreasing $T$ in the hadronic phase \cite{gavin} (because the relevant hadronic cross section decreases with $T$) while asymptotic freedom dictates that $\eta/s$ increases with $T$ in the deconfined phase since in this case the coupling between the quarks and the gluons (and the transport cross section) descreases logarithmically \cite{amy}. Note, however, that in general perturbative calculations are not reliable close to $T_c$ (see, however, Ref.\ \cite{Hidaka:2008dr}).

Thus far, there have been several attempts to compute $\eta/s$ in the hadronic phase using hadrons and resonances \cite{Itakura:2007mx,Gorenstein:2007mw,otheretasHG}. However, these studies have not explicitly considered that the hadronic density of states in QCD is expected to be $\sim \exp(m/T_H)$ for sufficiently large $m$ \cite{Hagedorn:1968jf,exponentialspectrum}, where $T_H \sim 150-200$ MeV is the Hagedorn temperature \cite{Hagedorn:1968jf} (see \cite{hagedornexperiment} for an update on the experimental verification of this asymptotic behavior). This hypothesis was originally devised to explain the fact that an increase in energy in $pp$ and $p\bar{p}$ collisions does not lead to an increase in the average momentum per particle but rather to production of more particles of different species \cite{Hagedorn:1968jf}. Moreover, hadron resonance models that include such rapidly increasing density of states are known to have a ``limiting" temperature, $T_{max}$, beyond which ordinary hadronic matter cannot exist \cite{Hagedorn:1968jf}.

In this letter, a hadron resonance gas model that includes all known particles and resonances with masses $m<2$ GeV \cite{PDG} and also an exponentially increasing number of Hagedorn states (HS) \cite{Noronha-Hostler,NHlong} is used to provide an upper limit on $\eta/s$ for hadronic matter close to $T_c$ that is comparable to $1/4\pi$. Additionally, we show that our model provides a good description of the recent lattice results \cite{Cheng:2007jq} for the trace anomaly and also the speed of sound, $c_s$, close to $T_c=196$ MeV. We comment on the effects of including HS on the bulk viscosity to entropy density ratio, $\zeta/s$, of hadronic matter near $T_c$.

The assumption behind hadron resonance models is the description of thermodynamic properties of a hadronic interacting gas by a free gas with these hadrons and their respective resonances. In \cite{welke} the pressure of an interacting pion gas calculated within the virial expansion nearly coincides with that of a free gas of pions and $\rho$ mesons. There is nearly exact cancellation between the attractive and repulsive S-wave channels \cite{prakash}. We assume that attractive interactions can be described by the inclusion of resonances which for large masses follow a Hagedorn spectrum. The system's mass spectrum is assumed \cite{Noronha-Hostler,NHlong} to be a sum over discrete and continuous states $\rho(m)=\rho_{HG}(m)+\rho_{HS}(m)$, where $\rho_{HG}(m)=\sum_{i}^{M_0}g_i\,\delta(m-m_i)\,\theta(M_0-m)$ involves a sum over all the known hadrons \cite{PDG} and their respective degeneracy up to $M_0<2$ GeV \cite{StatModel} and for larger masses
\begin{equation}
\ \rho_{HS}(m)=A\,\frac{e^{\,m/T_{H}}}{\left(m^2 +m_{0}^2\right)^{\frac{5}{4}}},
\end{equation}
where we take $m_{0}=0.5$ GeV, $A=0.5\;\textrm{GeV}^{\frac{3}{2}}$ \cite{Noronha-Hostler}, and $T_{H}=T_{c}$. In general, repulsive interactions among the hadrons soften the dependence of the pressure on the temperature \cite{prakash,hagedornrafelski,kapustavolcorr}. Their effects are included using the excluded-volume approach derived in \cite{kapustavolcorr} where the volume excluded by a hadron equals its energy divided by $4B$, where $B$ plays the role of an effective MIT bag constant. The thermodynamic quantities were found using
\begin{equation}
\ P(T)=\frac{P_{pt}(T^*)}{1-\frac{P_{pt}(T^*)}{4B}}\,,\qquad T=\frac{T^*}{1-\frac{P_{pt}(T^*)}{4B}}
\label{volumecorrections}
\end{equation}
and the standard thermodynamic identities at zero baryon chemical potential \cite{kapustavolcorr}. Note that the temperature $T$ and the pressure $P(T)$ of the system (after volume corrections) are defined in terms of the quantities computed in the point particle (subscript pt) approximation (i.e., no volume corrections). When $P_{pt}(T_c)/4B < 1$ there is still a limiting temperature that is larger than $T_c$ \cite{kapustavolcorr}. We take $B^{1/4}=0.34$ GeV in our calculations, which implies that $T_{max}>T_c$. We restrict our discussion to $T \leq T_c$ because at higher temperatures a description involving quarks and gluons should be more adequate.
\begin{figure}[t]
\centering
\epsfig{file=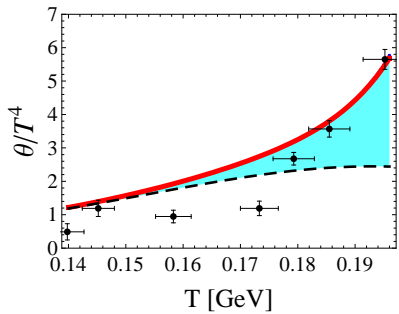,width=0.6\linewidth,clip=}
\caption{Comparison between $\theta(T)/T^4=\left(\epsilon-3P\right)/T^4$ using our hadron resonance gas model including Hagedorn states with $2<m<20$ GeV \cite{Noronha-Hostler} (solid red line) and our hadron gas model with only known hadrons up to $m<2$ GeV (black dashed line). The blue band between the curves illustrates the effects of HS. Repulsive interactions are included via an excluded volume approach \cite{kapustavolcorr} with $B^{1/4}=0.34$ GeV. Lattice data for the $p4$ action with $N_{\tau}=6$ \cite{Cheng:2007jq} is also shown.} \label{fig:trace}
\end{figure}

\begin{figure}[t]
\centering
\epsfig{file=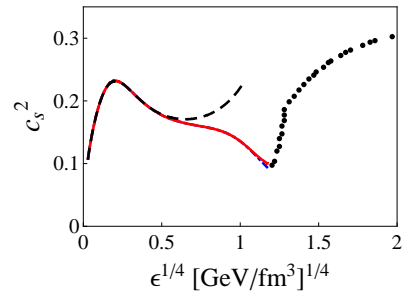,width=0.6\linewidth,clip=}
\caption{$c_s^2$ including HS with $2<m<20$ GeV (solid red line), HS with $2<m<80$ GeV (dotted-dashed blue), for a hadron gas model without HS (dashed black curve), and for $p4$ action lattice results with $N_{\tau}=6$ \cite{Cheng:2007jq} (dotted curve).} \label{fig:c_s}
\end{figure}

Our results for the trace anomaly are shown in Fig.\ \ref{fig:trace} where the mass of the heavier Hagedorn state was set to be $M_{max}=20$ GeV. Note that the inclusion of HS correctly captures the trend displayed by the lattice data in the transition region whereas our hadron gas curve does not \cite{comment}. This remains true if other values of $B$ are used. We checked that our results did not change appreciably in this temperature range when $M_{max}$ is increased to $80$ GeV. This happens because the divergences normally associated with the limiting temperature only occur in this case at $T_{max}\sim 210$ MeV. Were $T_H < T_c$, the dependence of the thermodynamic quantities with $M_{max}$ would be much more pronounced. In general, a very rapid increase in the number of particle species (specifically heavier species) around $T_c$ is expected to strongly reduce the speed of sound $c_s^2=dP/d\epsilon$ at the phase transition. While $c_s^2 \to 0$ at the transition would certainly lead to very interesting consequences for the evolution of the RHIC plasma \cite{Rischke:1996em}, recent lattice simulations have found that $c_s^2\simeq 0.09$ near $T_c$ \cite{Cheng:2007jq}. It is shown in Fig.\ \ref{fig:c_s} that $c_s^2(T\sim T_c)\sim 0.09$ in the model with HS while for the model without them $c_s^2 \sim 0.25$ near the transition. Note that when $M_{max}=80$ GeV (dashed blue curve) $c_s^2$ is only a bit smaller than $0.09$ near the phase transition. Other quantities such as the total entropy density near $T_c$ are found to agree with lattice results within the uncertainties present in those calculations \cite{Bazavov:2009zn}.


The total shear viscosity of our multi-component system computed within kinetic theory \cite{Reif} is $\eta_{tot} \sim \alpha \sum_i n_i \langle p_i \rangle \lambda_i$, where $n_{i}$ is the number density, $\langle p_{i} \rangle$ is the average momentum, and $\lambda_{i}$ is the mean free path for discrete states and HS ($\alpha\sim \mathcal{O}(1)$). Moreover, $\lambda_{i}=\left(\sum_{j} n_j\, \sigma_{ij}\right)^{-1}$ where $\sigma_{ij}$ is the scattering cross section. Due to their large mass, the particle density of HS is much smaller than that of discrete states. Thus, one can neglect the small contribution to the mean free path from terms involving the interaction between the standard hadrons and the HS. In this case, $\eta_{tot}=\eta_{HG}+\eta_{HS}$ where $\eta_{HG}$ is the shear viscosity computed using only the interactions between the standard hadrons while $\eta_{HS}=\frac{1}{3}\sum_{i}n_i \langle p\rangle_i \,\lambda_{i}$ includes only the contribution from HS, which move non-relativistically since $m_{HS}/T\gg 1$. Note that our approximation for $\eta_{tot}$ provides an upper bound since the inclusion of the interactions between HS and hadrons would decrease $\eta_{tot}$. Using the results above, 
\begin{eqnarray}
\label{etas1}
\left(\frac{\eta}{s}\right)_{tot}&\leq&\frac{s_{HG}}{s_{HG}+s_{HS}}\left[\left(\frac{\eta}{s}\right)_{HG}+ \frac{\eta_{HS}}{s_{HG}}\right].
\end{eqnarray}
While the entropy dependent prefactor in Eq.\ (\ref{etas1}) can be easily determined using our model, the detailed calculation of $\eta_{HG}$ and $\eta_{HS}$ requires the knowledge about the mean free paths of the different particles and resonances in the thermal medium. In the non-relativistic approximation, we can set $\langle p_i \rangle = m_i \langle v_i \rangle = \sqrt{8m_i\,T/\pi}$ in Eq.\ (\ref{etas1}). Note that HS with very large $m_i$'s are more likely to quickly decay. We assume that $\lambda_i=\tau_i \,\langle v_i\rangle$ where $\tau_i\equiv 1/\Gamma_i=1/(0.151\, m_i-0.0583)$ GeV$^{-1}$ is the inverse of the decay width of the $i^{th}$ HS obtained from a linear fit to the decay widths of the known resonances in the particle data book \cite{Noronha-Hostler,NHlong,Senda}. Our choice for $\lambda_i$ gives the largest mean free path associated with a given state because it neglects any possible collisions that could occur before it decays on its own. Note, however, that the decay cross section is in general different than the relevant collision cross section for momentum transport that contributes to $\eta$ according to kinetic theory. Thus, it is not guaranteed a priori that these decay processes contribute to $\eta$ in the usual way. Further studies of the relationship between HS and $\eta$ could be done, perhaps, using the cross sections discussed in \cite{Pal:2005rb}.

\begin{figure}[t]
\centering
\epsfig{file=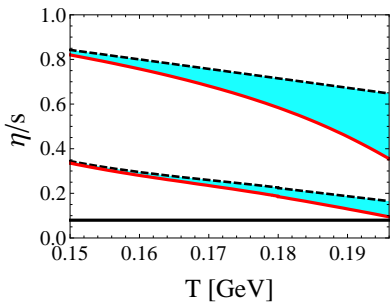,width=0.6\linewidth,clip=}
\caption{$\eta/s$ is shown for a gas of $\pi$'s and nucleons \cite{Itakura:2007mx} (upper dashed black line), for a hadron resonance gas with excluded volume corrections \cite{Gorenstein:2007mw} (lower dashed black line), and for KSS limit, $\eta/s=1/4\pi$, \cite{Kovtun:2004de} (solid black). An upper bound from HS on $\eta/s$ (solid red) and the effects of HS (blue band) are shown.} \label{fig:eta_s}
\end{figure}

Substituting the results above in we find that $\eta_{HS} =8T\sum_{i}\,n_{i}\tau_i/3\pi$. The remaining ratio $\left(\eta/s\right)_{HG}$ has been computed in Refs.\ \cite{Itakura:2007mx,Gorenstein:2007mw,otheretasHG} using different models and approximations. Since our main goal is to understand the effects of HS on  $(\eta/s)_{tot}$, here we will simply use the values for $\left(\eta/s\right)_{HG}$ obtained in some of these calculations to illustrate the importance of HS. We chose to obtain $(\eta/s)_{HG}$ for a gas of pions and nucleons from Fig.\ 5 in \cite{Itakura:2007mx} and for a hadron resonance gas with (constant) excluded volume corrections from \cite{Gorenstein:2007mw}. Note that the results for $\eta/s$ obtained from the calculation that included many particles and resonances \cite{Gorenstein:2007mw} are already much smaller than those found in \cite{Itakura:2007mx} where only pions and nucleons are considered. A linear extrapolation of the results in \cite{Itakura:2007mx,Gorenstein:2007mw} was used to obtain their $\eta/s$ values at high temperatures. In Fig.\ \ref{fig:eta_s}, $\left(\eta/s\right)_{tot}$ drops significantly around $T_{c}$ because of HS.  This result is especially interesting because $\eta/s$ in the hadronic phase is generally thought to be a few times larger than the string theory bound. One can see that the contributions from HS should lower $\eta/s$ near to the KSS bound close to $T_c$. Thus, the drop in $\eta/s$ due to HS could explain the low shear viscosity near $T_c$ already in the hadronic phase. We used $M_{max}=20$ GeV in Fig.\ \ref{fig:eta_s} but the results do not change significantly if $M_{max}$ is increased by a factor of 4.

\begin{figure}[t]
\centering
\epsfig{file=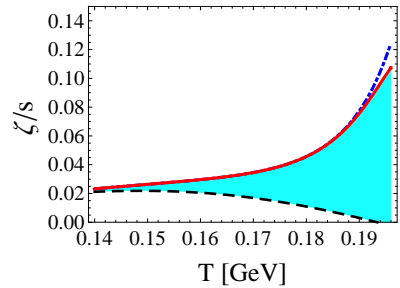,width=0.6\linewidth,clip=}
\caption{Estimates for $\zeta/s\equiv G^{E}(0)/(9\omega_0 \,s)$ ($\omega_0=1$ GeV) for the model that includes HS with $2<m<20$ GeV (solid red line) and $2<m<80$ GeV (dotted-dashed blue line) and our hadron gas model with $m<2$ GeV (black dashed line).} \label{fig:bulk}
\end{figure}

The large value of the trace anomaly near $T_c$ observed on the lattice has been used as an indication that $\zeta/s$ of QCD may be large at the phase transition \cite{Meyer:2007dy,dima}. This is very different than at high temperatures where $\zeta/s$ is predicted to be small \cite{hosoya}. This may have some interesting phenomenological consequences such as the formation of clusters at freeze-out \cite{Torrieri:2007fb}. Using QCD sum rules in \cite{Ellis:1998kj}, one can extract the (zero-momentum) Euclidean correlator of the energy-momentum tensor's trace, $\theta^{\mu}_{\mu}$:
\begin{eqnarray}\label{zetaDima}
G^{E}(0,{\bf 0})&=&\int d^4 x\,\langle \theta^{\mu}_{\mu}(\tau,{\bf x})\theta^{\nu}_{\mu}(0,{\bf 0})\rangle \nonumber \\ &=&\left(T\partial_T -4\right)\left(\epsilon-3p\right)
\end{eqnarray}
According to  \cite{dima}, $\zeta$ can be obtained via $G^{E}$ using $\pi\rho(\omega,{\bf 0})/9\omega=\zeta \omega_0^2/(\omega^2+\omega^2_0)$ as an ansatz for the small frequency limit of the $\langle \theta\theta \rangle$ spectral density at zero spatial momentum, $\rho(\omega,{\bf 0})$. The parameter $\omega_0(T)$ defines the energy scale at which perturbation theory is applicable. The validity of this ansatz has been recently studied in Refs.\ \cite{validityDima}. Here we assume that this ansatz can at least capture the qualitative behavior of $\zeta$ around $T_c$ and we use it to estimate how HS change the $\zeta/s$ close to $T_c$. The results for $\zeta/s\equiv G^{E}(0)/(9\omega_0 \,s)$ are shown in Fig.\ \ref{fig:bulk} where $\omega_0=1$ GeV. While $\zeta/s$ decreases near $T_c$ for the hadron gas model, $\zeta/s$ including HS increases close to $T_c$ and this enhancement does not vary much with $M_{max}$.

In conclusion, a hadron resonance gas model including all the known particles and resonances with masses $m<2$ GeV and also an exponentially rising level density of Hagedorn states for $m>2$ GeV was used to obtain an upper bound on $\eta/s$ for hadronic matter near $T_c$ that is comparable to the KSS bound $1/(4\pi)$. The large trace anomaly and the small $c_s$ near $T_c$ computed within this model agree well with recent lattice calculations \cite{Cheng:2007jq}. Moreover, according to the general result that small $\eta/s$ implies strong jet quenching \cite{Majumder:2007zh}, our significant reduction of $\eta/s$ indicates that hadronic matter near the phase transition is more opaque to jets than previously thought. 

We thank G.\ Torrieri, K.\ Redlich, S.\ Bass, and A.\ Dumitru for interesting discussions. J.N. acknowledges support from US-DOE Nuclear Science Grant No. DE-FG02-93ER40764. This work was supported by the Helmholtz International Center for FAIR within the LOEWE program launched by the State of Hesse.



\begin{thebibliography}{99}
\bibitem{comment} Note that the hadron gas curve displayed in Fig.\ 1 is different than that in \cite{Cheng:2007jq} because we only include states with $m<2$ GeV and the corresponding trace anomaly is also reduced due to the excluded-volume corrections.
\end{thebibliography}
\end{document}